\documentclass[aps,pre,eqsecnum,twocolumn]{revtex4}
\usepackage{amsmath,bm,epsfig}

\def\Fbox#1{\vskip1ex\hbox to 8.5cm{\hfil\fboxsep0.3cm\fbox{%
  \parbox{8.0cm}{#1}}\hfil}\vskip1ex\noindent}  
\let \nn  \nonumber
\let \ve \varepsilon
\newcommand{\br}{\\ \nn}
\def\r{\bm r}

\def\s{\bm s}
\def\S{\bm S}
\def\v{\bm v}
\def\V{\bm V}

\let\*\cdot
\def\<{\left\langle} \def\>{\right\rangle} \def\({\left(} \def\){\right)}
\let\p\partial \let\~\widetilde \let\^\widehat \def\ort#1{\^{\bf{#1}}}
\def\Trans{^{\scriptscriptstyle{\rm T}}} \def\x{\ort x} \def\y{\ort y}
\def\z{\ort z} \def\bn{\bm\nabla} \def\1{\bm1} \def\Tr{{\rm Tr}}
\newcommand{\B}[1]{{\bm{#1}}}
\newcommand{\C}[1]{{\mathcal{#1}}}    
\newcommand{\BC}[1]{\bm{\mathcal{#1}}}

\renewcommand{\sb}[1]{_{\text {#1}}}  

\newcommand{\Ref}[1]{(\ref{#1})}
\newcommand{\REF}[1]{Eq.~(\ref{#1})}

\newcommand{\g}{\gamma}
\renewcommand{\d}{\delta}

\renewcommand{\t}{\tau}
\let \= \equiv

\def\nn{\nonumber}

\def\V{\bm V} \def\r{\bm r} \let\p\partial \def\bn{\bm\nabla}
 \let\*\cdot \def\sb#1{_{\rm{#1}}}
 
 \def\({\left(} \def\){\right)}
 \def \[ {\left [} \def \] {\right ]}
  \def\S{\bm S} \def\Sp#1{^{\scriptscriptstyle\rm{#1}}}
  \def\1{\bm\delta} \def\v{\bm v} \let\^\widehat \def\REF#1{Eq.~\Ref{#1}}
   \def\Ref#1{(\ref{#1})} \def\k{{\bm k}} \def\<{\left\langle}
   \def\>{\right\rangle} \def\s{\bm s} \def\x{\ort x}
   \def\ort#1{\^{\bf{#1}}} \def\y{\ort y} \def\z{\ort z}
     \def\Sb#1{_{\scriptscriptstyle\rm{#1}}}
     \let\~\widetilde
\def\Re{\ensuremath{{\C R}\mkern-3.1mu e}}
\def\De{\ensuremath{{\C D}\mkern-3.1mu e}}

\begin{document}

\title{The Polymer Stress Tensor in Turbulent Shear Flows}
\author{Victor S. L'vov, Anna Pomyalov, Itamar Procaccia
and Vasil Tiberkevich}

\affiliation{~Department of Chemical Physics, The Weizmann
Institute of Science,  Rehovot 76100, Israel}
\begin{abstract}

The interaction  of polymers with  turbulent shear flows is
examined. We focus on the structure of the elastic stress tensor,
which is proportional to the polymer conformation tensor. We
examine this object in turbulent flows of increasing complexity.
First is isotropic turbulence, then anisotropic (but homogenous)
shear turbulence and finally wall bounded turbulence. The main
result of this paper is that for all these flows the polymer
stress tensor attains a universal structure in the limit of large
Deborah number $\De\gg 1$. We present analytic results for the
suppression of the coil-stretch transition at large Deborah
numbers. Above the transition the turbulent velocity fluctuations
are strongly correlated with the polymer's elongation: there
appear high-quality ``hydro-elastic" waves in which turbulent
kinetic energy turns into polymer potential energy and vice versa.
These waves determine the trace of the elastic stress tensor but
practically do not modify its universal structure. We demonstrate
that the influence of the polymers on the balance of energy and
momentum can be accurately described by an effective polymer
viscosity that is proportional to to the cross-stream component of
the elastic stress tensor. This component is smaller than the
stream-wise component by a factor proportional to $\De ^2 $.
Finally we tie our results to wall bounded turbulence and clarify
some puzzling facts observed in the problem of drag reduction by
polymers.
\end{abstract}
\maketitle

\section{Introduction}\label{s:intro}

The dynamics of dilute polymers in turbulent flows is a rich
subject, combining the complexities of polymer physics and of
turbulence. Besides fundamental questions there is a significant
practical interest in the subject, particularly because of the
dramatic effect of drag reduction in wall bounded turbulent
flows~\cite{49Tom,75Vir,90Gen,00SW}. On the one  hand the polymer
additives provide a channel of dissipation in addition to the
Newtonian viscosity; this had been stressed in the past mainly by
Lumley~\cite{69Lum,72Lum}. On the other hand the polymers can
store energy in the form of elastic energy; this aspect had been
stressed for example by Tabor and De Gennes~\cite{86TG}. A better
understanding of the relative roles of these aspects requires a
detailed analysis of the dynamics of the ``complex fluid" obtained
with dilute polymers in the turbulent flow regime.

Some important progress in the theoretical description of the
statistics of polymer stretching in homogeneous, isotropic
turbulence of dilute polymer solutions was offered in
Refs.~\cite{00Che,00BFL,01BFL}. Here we want to stress the fact
that in the practically interesting turbulent regimes, and in
particular when there exists a large drag reduction effect, the
characteristic mean velocity gradient (say, the mean shear,
$S_0$), is much larger than the inverse polymer relaxation time,
$\tau\sb p$, $S_0\tau\sb p\gg 1$. Usually this parameter is
referred to as the Deborah or Weissenberg number:
\begin{equation}\label{De}
\De\equiv \C W\! i \equiv S_0\tau\sb p\ .
\end{equation}%
 Indeed, the onset of drag reduction corresponds to Reynolds
number $\Re\sb {cr}$ at which $\De \sim 1$. Large drag reduction
corresponds to $\Re \gg \Re \sb {cr}$ at which $\De\simeq \Re /\Re
\sb{cr}\gg 1$.

The aim of this paper is to provide a theory of the polymer stress
tensor $\bm{\Pi}$ [defined in Eq. (\ref{def-Pi})] in turbulent flows in which $\De \gg 1$. The
main result is a relatively simple form~\Ref{Pi-a} of
the mean polymer stress tensor $\bm{\Pi}_0$  that enjoys a high degree
of universality. Denoting by  $\x$ and $\y$ the unit vectors in the
mean velocity and mean velocity gradient directions (streamwise
and cross-stream directions in a channel geometry) and by
$\z=\x\times \y$ the span-wise direction,  we show that tensor
$\bm{\Pi}_0$ has a universal form:
\begin{subequations}\label{Pi}\begin{equation}
    \bm\Pi_0 \simeq \Pi^{yy}_0\(\begin{array}{ccc}
        2\De ^ 2 & ~\De~ &  ~~0~~\\
        \De & 1 & 0 \\
        0 & 0 & C
    \end{array}\) \quad \mbox{for}\ \De\gg 1\ .\label{Pi-a}
    \end{equation}
One sees that in the ($\x$-$\y$)-plane the tensorial structure of $
\bm\Pi_0$ is  independent of the statistics of turbulence:
\begin{equation}\label{Pi-b}
\Pi^{xx}_0=2\, \De^2 \Pi^{yy}_0\,, \qquad
\Pi^{xy}_0=\Pi^{yx}_0=\De \,\Pi^{yy}_0\ .
\end{equation}%
Due to the to the symmetry of the  considered flows with respect
to reflection $z\to -z$, the off-diagonal components $\Pi_0^{xz}$
and $\Pi_0^{yz}$ vanish identically:
\begin{equation}\label{Pi-c}
\Pi_0^{xz}=\Pi_0^{yz}= 0\ . 
\end{equation}%
\end{subequations}%
The only non-universal entry in \REF{Pi-a} is the dimensionless coefficient $C$. We show that $C=1$ for  
shear flows in which the extension of the polymers is caused by temperature fluctuations and/or by
isotropic turbulence. For anisotropic turbulence the constant $C$ is
of the order of unity.

The universal tensorial structure \Ref{Pi} has important
consequences for the problem of drag reduction in dilute polymeric
solutions. We show that the effect of the polymers on the balance
of energy and mechanical momentum can be described by an effective
polymer viscosity that is proportional to the cross-stream
component of the elastic stress tensor $\Pi_0^{yy}$. According to
\REF{Pi-b} this component is smaller than the stream-wise
component $\Pi_0^{xx}$ by a factor of $2\, \De ^2 $. This  finding
resolves the puzzle of the linear increase of the effective
polymeric viscosity with the distance from the wall, while the
total elongation of the polymers [dominated by the largest
component $\Pi^{xx}_0$ of the tensor \Ref{Pi-a}] decreases with
the distance from the wall.  We show that our results for the
profiles of the components $\Pi ^{ij}_0$ are in a good agreement
with DNS of the FENE-P model in turbulent channel flows.

 The paper is organized as follows. In Sec.~\ref{ss:BEMa} we
present the basic equations of motion of the problem.  In
Sec.~\ref{ss:BEMb} we describe  the standard Reynolds
decomposition of the equations of motion for the mean and
fluctuations of the relevant variables. In Sec.~\ref{ss:BEM-c} we
find general solution  for $\bm{\Pi}_0$ and formulate (the very weak)
conditions at which it simplifies to the universal form \Ref{Pi}.

In Section~\ref{s:PS} we analyze the phenomenon of polymer
stretching in shear flows with prescribed turbulent velocity
fields, either isotropic, Sec.~\ref{ss:PSb}, or anisotropic,
Sec.~\ref{ss:PSc}. We relate the value of $C$ to the level of anisotropy
in the turbulent flow.   We also show in this section that a
strong mean shear suppresses the threshold value for the
coil-stretched transition by a factor $\De ^ 2$, see
Eqs.~\Ref{th}.

Sections~\ref{s:Act} and~\ref{s:Disc}  are devoted to the case of
strong turbulent fluctuations for which the fluctuating velocity
and polymeric fields are strongly correlated. In
Sec.~\ref{s:Act} we consider the case of homogeneous shear
flows.  In Sec.~\ref{s:Disc} we discuss  wall bounded
turbulence in a channel geometry, compare our results with
available DNS data and apply  our finding to the problem of drag
reduction by polymeric solutions.

Section~\ref{s:sum} presents summary and a
discussion of the obtained results .

Appendix~\ref{ss:BEMc} offers some useful  exact relationships
for the energy balance in the system.

\section{Equations of Motion and Solution for Simple Flows}
\label{s:BEM}

\subsection{Equations of motion for dilute polymer solutions}\label{ss:BEMa}

\paragraph{The equation for the velocity field of the complex fluid} reads:
\begin{subequations}\label{NSE}
\begin{eqnarray}\nn
\frac{\p\,  \V}{\p t} +\(\V\*\bn\)\V &=& \nu_0\Delta\V -\bn P
+\bn\*\bm\Pi \,,\nonumber \\
\bn \*\V &=& 0 \ . \label{NSEa} 
\end{eqnarray} 
Here $\V\equiv\V(t,\r)$, $P\equiv P(t,\r)$ and $\nu_0$ are the fluid
velocity,
the pressure, and Newtonian viscosity of the neat fluid respectively. The
fluid is considered incompressible, and units are chosen such that the
density
is 1.
The effect of the added polymer appears in \REF{NSEa} via
the elastic stress tensor $\bm\Pi\equiv\bm\Pi(t,\r)$.

In this paper we consider dilute polymers in the limit that their
extension due to interaction with the fluid is small compared with
their chemical length. In this regime one can safely assume that
the  polymer extension is proportional to the applied force
(Hook's law). We simplify the description of the polymer dynamics
assuming that the relaxation to equilibrium is characterized by a
single relaxation time $\tau\sb p$, which is constant at small
extensions.  In this case $\bm\Pi$ can be written as
    \begin{equation}\label{def-Pi}
    \bm\Pi(t,\B r)\equiv \frac{\nu_{0\rm p}}{\tau\sb p}\overline{\bm R(t,\B r)\bm R(t,\B r)}\,,
    \end{equation}
where $\nu_{0\rm p}$ is polymeric viscosity for infinitesimal
shear, and $\bm R$ is the polymer end-to-end elongation vector,
normalized by its equilibrium value (in equilibrium $\overline{\bm R\bm
R}=\1$). The average in \REF{def-Pi} is over the Gaussian statistics of the
Langevin random force which describes the interaction of the polymer molecules
with the solvent molecules at a given temprature, but not over the turbulent ensemble. 

\paragraph{The equation for the elastic stress tensor $\bm\Pi$}
has the form 
\begin{eqnarray} \frac{\p\,  \B\Pi}{\p t} +(\V\*\bn){\B\Pi} &=&
\S\*{\B\Pi} +{
\B\Pi}\*\S\Trans -\frac{1}{\tau\sb p}\({\B\Pi} -\B\Pi\sb{eq}\)
\,,\nonumber\\
  \S\equiv\(\bn\V\)\Trans,&& \quad  S^{ij}=\p V^i/\p
x^j\,,\label{NSEb}\\\nn 
\bm{\Pi}\sb{eq}= \Pi\sb{eq}\1\,, &&  \quad
\Pi\sb{eq}\equiv\nu_{0\rm p}/\tau \sb p\ .
\end{eqnarray}
\end{subequations}
Here $\S\equiv\S(t,\r)$ is the velocity gradient tensor and $\1$ is
unit tensor.

\paragraph{Choice of coordinates.}
In this paper we consider both homogeneous and wall bounded shear
flows. In both cases we choose the coordinates such that the mean
velocity is in the $\x$ direction and the gradient is in the $\y$
direction. In wall bounded flows $\x$ and $\y$ are unit vectors in
the streamwise and the wall normal directions:
\begin{eqnarray}\label{R-dec-a}\label{R-dec}
\V_0 &\equiv& \<V\> = \S_0\*\r \,,\quad \S_0 \equiv S_0\,\x\y
\,,\br
 && \text{only}\   V_0^x=S_0 \, y \ne 0\  \text{and} 
 \ S_0^{xy}=S_0\ne 0\ .\end{eqnarray} 
 The  unit vector orthogonal
to $\x$ and $\y$ (the span-wise direction in wall bounded cases)
is denoted as $\z$.

\subsection{Reynolds decomposition of the basic equations}\label{ss:BEMb}

In the following we need to consider separately the mean values
and the fluctuating parts of the velocity, $\B V(\r,t)$, the
velocity gradient $\B S(\r,t)$,  and the elastic stress tensor
$\bm{\Pi}(\r,t)$ fields. We define the fluctuating parts via
\begin{equation}\label{R-dec-b}
\V=\V_0+\v\,, \quad \S=\S_0+ \s\,, \quad {\B\Pi}={\B\Pi}_0+\bm
\pi\,,\dots
\end{equation}
All the mean quantities will be denoted with a ``$_0$'' subscript
(e.g. the mean velocity $\V_0$), and all the fluctuating
quantities by the lower-case letters $\v$, $p$, $\bm\pi$, $\s$,
etc. These mean values are computed with respect to the appropriate turbulent ensemble. Note that the mean
pressure in a homogeneous shear flow is zero, $P_0=0$, and all mean quantities (except the mean velocity
$V_0^x = S_0\,  y$, of course) are coordinate independent and, by
definition, time independent.

\subsubsection{Equations for the mean objects}
\paragraph{The equation for the mean elastic stress tensor} $\bm{\Pi}_0$
follows
from  \REF{NSEb}:
\begin{subequations}\label{eqR0}
 \begin{eqnarray} \label{eqR0a}
&&\hskip -2.1cm  \(\frac{D}{D\, t} +\frac{1}{\tau\sb
p}\){\bm{\Pi}}_0 = \S_0\*{\bm{\Pi}}_0 +{\bm{\Pi}}_0\*\S_0\Trans
+\bm{Q}\,,
\\ \label{eqR0b}
&&   \bm{Q} \equiv \frac{1}{\tau\sb p}\, \bm{\Pi}\sb {eq}+
\bm{Q}\Sb T \,,
   \\  \label{eqR0c}&&\hskip -0.15cm
  \bm{Q}\Sb T \equiv  \<\s\*\bm\pi +\bm\pi\*\s\Trans\>\,,
\end{eqnarray} 
where $\bm{\Pi}\sb {eq}$ is given by \REF{NSEb}, and $\langle\cdots\rangle$ denotes an
average with respect to the turbulent ensemble.  In components:
\begin{eqnarray}\nn
&& \Big(\frac{D}{D\, t}+ \frac{1}{\tau\sb p}\, \Big )\Pi_0^{ij} =
S_0^{ik}\Pi_0^{kj}+S_0^{jk}\Pi_0^{ki}+ Q^{ij}\,,
\\
 && Q^{ij}=\frac1{\tau\sb p}\Pi
\sb{eq}\d^{ij} + \< s^{ik}\pi ^{kj}+s^{jk}\pi ^{ki} \> \ .
\end{eqnarray}  \end{subequations}
Here $D/D t$ is the mean substantial time derivative 
\begin{equation} \frac{D}{D\, t} \equiv \frac{\p}{\p t} +\V_0\*\bn =
\frac{\p}{\p t} + S_0\, y \, \frac{\p}{\p x}\ .
\end{equation}
The substantial derivative vanishes in the stationary case, when
all the statistical objects are $t$- and $x$-independent.

\paragraph{The equation for the mean velocity} follows from \REF{NSEa}:
 \begin{equation}\label{flux-a}
 \frac{\partial V_0}{\partial t}=\frac{\partial }{\partial y}
 \left[ \nu_0S_0 -W^{xy} +\Pi_0^{xy} \right]\,,
\end{equation}%
were $\B W$ is the Reynolds stress tensor:
\begin{equation}\label{RS} 
W^{ij} \equiv \< v^iv^j\>\ . \end{equation} 
In the stationary case $\partial V_0/\partial t=0$ and
\REF{flux-a} gives
\begin{equation}\label{bal1}
\nu_0S_0 -W^{xy} +\Pi_0^{xy}= {\cal P}\,,
\end{equation}
where  constant of integration $\cal P$ has a physical meaning of
the total momentum flux. The three terms on the LHS of \REF{bal1}
describe the viscous, inertial and polymeric contributions to
$\cal P$. In a homogeneous-shear geometry all theses terms are
$y$-independent constants, whereas in a channel geometry they depend on
the distance to the wall. Notice, that \REF{bal1} also can be
considered as an equation for the balance of mechanical forces in
the flow.

\subsubsection{Equations for the fluctuations}
The equations for the fluctuating parts $\v$ and $\bm \pi$ read:\
\begin{subequations}\label{15}
\begin{eqnarray}
&&\frac{D\, \v }{D\, t} = -\S_0\*\v +\nu_0\Delta\v -\bn p
+\bn\*\(\bm\pi -\v\v\) \,,\nonumber \\
\label{15a}\\
 && \(\frac{D}{D\, t}
+\frac{1}{\tau\sb p}\)\bm\pi = \S_0\*\bm\pi
+\bm\pi\*\S_0\Trans+\s\*{\bm{\Pi}}_0 +{\bm{\Pi}}_0\*\s\Trans  \nonumber  \\
\label{15b} &&~~~~~~~~~~~~~~~~~~~~~~~~~~~~~~~~~
 +\bm\Phi_\pi
\,,\end{eqnarray} where $\bn\*\v = 0 $,  and
\begin{eqnarray}
\bm\Phi_\pi &\equiv& \(\s\*\bm\pi +\bm\pi\*\s\Trans\)^{\~{}}
-\bn\*\v\bm\pi \ .\end{eqnarray} \end{subequations}
Here $(\dots)^{\~{}}$ denotes the ``fluctuating part of" $(\dots)$.

\subsection{Solutions for simple flow configurations}
\label{ss:BEM-c} 
\subsubsection{Implicit solution  for shear flows}

Note that the mean shear tensor $\bm S_0$ satisfies, besides the
incompressibility  condition ${\rm Tr}\,\{\bm S_0\}=S_0^{ii}=0$,
one additional constraint
\begin{equation}
S^{ij}_{0}S^{jk}_{0}=0 \ . \label{simple}
\end{equation}
Remarkably, this property of $\bm S_0$ uniquely distinguishes shear flows from
other possibilities (elongational or rotational flows): if \Ref{simple} holds,
one can always choose coordinates such that the only nonzero
component of $\bm{S}_{0}$ is $S_0^{xy}=S_0$.

 Having \REF{simple} consider \REF{eqR0a} in the stationary state:
\begin{equation}\label{Ro-a}
 {\bm{\Pi}}_0 = \tau\sb p\(\S_0\*{\bm{\Pi}}_0
   +{\bm{\Pi}}_0\*\S_0\Trans + \bm{Q}\)\ .
\end{equation}%
We can proceed to solve this equation implicitly, treating $\bm{Q}$
on the RHS as a given tensor, and solve the linear set of equation
for $\bm{\Pi}_0$. In the considered geometry \REF{Ro-a} is a
system of 4 linear equations, and the solution is expected to be
quite cumbersome. However, property \Ref{simple} allows a very
elegant solution of this system. Using \Ref{Ro-a} iteratively
(i.e., substituting instead of $\bm{\Pi}_0$ on the RHS of
\REF{Ro-a} the whole RHS),  we get
\begin{eqnarray}\nonumber
  {\bm{\Pi}}_0 &=& 2\tau\sb p^2\S_0\*{\bm{\Pi}}_0\*\S_0\Trans
    +\tau\sb p^2\(\S_0\*\bm Q +\bm Q\*\S_0\Trans\)
    +\tau\sb p\bm{Q}
\\\nonumber&&
    +\tau\sb p^2\(\S_0^2\*\bm Q +\bm Q\*\(\S_0^2\)\Trans\)
\ .\end{eqnarray} One sees that due to \REF{simple} the last term
in this equation vanishes. Repeating this procedure once again, we
obtain the {\em exact} solution of \Ref{Ro-a} in the form of a
finite (quadratic) polynomial of the tensor $\bm{S}_0$:
\begin{subequations}\label{R0-sol}
\begin{equation}\label{R0-sol-a} 
\bm{\Pi}_0= 2\, \tau\sb p ^ 3 \, \S_0   \* \bm{Q} \* \S_0 \Trans +
\tau \sb p^ 2 \(\S_0\* \bm{Q} + \bm{Q} \* \S_0\Trans \)+ \tau\sb
p\bm{Q}\ .\end{equation}
The individual components of the solution~\REF{R0-sol-a} are given
by:
\begin{eqnarray} \nn 
\Pi^{xx}_0&=& \tau\sb p\(2\,\De ^ 2\,Q^{yy}+ 2\,\De\,
Q^{xy}+Q^{xx}\)\,,\\ \label{R0-sol-b}
\Pi^{xy}_0&=&  \tau\sb p\(\De\, Q^{yy}+Q^{xy}\)\,, \br 
\Pi^{yy}_0&=& \tau\sb pQ^{yy}\,,  \qquad \Pi^{zz}_0= \tau\sb
pQ^{zz}\,,\br 
\Pi_0^{xz}&=&\Pi_0^{yz}= 0\ . 
\end{eqnarray} 
\end{subequations}
 Notice that the ${xz}$ and $yz$ components of $\bm{\Pi}_0$ vanish due
to the symmetry of reflection $z\to -z$, which remains relevant in all the
flow configurations addressed in this paper.

In the limit $\De \gg 1$ the tensorial structure~\Ref{R0-sol-b} can be
simplified to the form \Ref{Pi}  if 
\begin{equation}\label{case} Q^{xy}\ll \De\,
Q^{yy}\,, \quad  Q^{xx}\ll \De ^ 2\,  Q^{yy}\ .
\end{equation}%
Indeed, in this case one can neglect the last two terms in
$\Pi_0^{xx}$ and the second term in $\Pi_0^{xy}$ and obtain
\REF{Pi} with $C=Q^{zz}/Q^{yy}$.

\subsubsection{Explicit solution for laminar homogeneous shear flows}
The solution \Ref{R0-sol} allows further simplification in the
case of  {\em laminar} shear flow, in which $\bm{Q}\Sb T$ of Eq.
(\ref{eqR0c}) vanishes, $\bm{Q}\Sb T=0$,  and thus,  according to
Eqs.~\Ref{eqR0b} and \Ref{NSEb},  the tensor $\bm{Q}$ becomes
proportional to the unit tensor:
 \begin{equation}\label{Q}
Q^{ij}=\frac{\Pi\sb{eq}}{\tau\sb p}\,\d^{ij}\ .
\end{equation}%
 With this relationship  \REF{R0-sol-a} simplifies to
    \begin{equation}
    \bm\Pi_0 = \Pi\sb{eq}\(\begin{array}{ccc}
        2\De ^ 2+1 & ~\De~ &  ~~0~~\\
        \De & 1 & 0 \\
        0 & 0 & 1
    \end{array}\)\ .\label{R0eq}
    \end{equation}

Equation  \Ref{R0eq}, in which   $\Pi\sb{eq}$ is given by
\REF{NSEb}, is important in itself as an \emph{explicit}
 solution for $\bm{\Pi}_0$ in the case of laminar shear flow. But more
importantly,  this equation together with \REF{R0-sol} gives  a
hint regarding the tensorial structure of  $\bm{\Pi}_0$ even in
the presence of turbulence, when the coupling between the velocity
and the polymeric elongation field, leading to the
cross-correlation $\bm{Q}\Sb T$, cannot be neglected.

To see why the simple result \Ref{R0eq} may be relevant also for
the turbulent case, note first that the non-diagonal elements
with $z$ projection, i.e. $\Pi_0^{xz}$ and $\Pi_0^{yz}$,  are
identically zero in general as long as the symmetry $z\to -z$ prevails.
Second, for large Deborah numbers, $\De
\gg 1$, the nonzero components in both  Eqs.~\Ref{R0eq} and
\Ref{R0-sol-b} have three different orders of magnitude: $\Pi_0^{xx}\gg
\Pi_0^{xy}\gg \Pi_0^{yy}\simeq \Pi ^{zz}$. In other words, the tensor
$\bm{\Pi}_0$ is strongly anisotropic, reflecting a strong
preferential orientation of the stretched polymers along the
stream-wise direction $\x$. The characteristic deviation angle
(from the $\x$ direction) is of order $O(1/\De)$.

Notice  that  Eqs.~\Ref{R0-sol} relate the polymeric stress tensor
$\bm{\Pi}_0$ to the cross-correlation tensor $\bm{Q}\Sb T$,
\REF{eqR0c}. In its turn this tensor depends on the polymeric
stretching, which is described by the same tensor $\bm{\Pi}_0$.
Therefore, generally speaking, Eqs. \Ref{R0-sol} remains an
\emph{implicit}  solution of the problem, which in
general requires considerable further analysis. However, as we
discussed above, \REF{R0-sol-a} is more transparent than the
starting \REF{eqR0}. In particular,  if the tensor $\bm{Q}$ is not
strongly anisotropic, the tensorial structure of $\bm{\Pi}_0$ is
close to \REF{R0eq} for the laminar case. We will see below that
this structure is indeed recovered under more general conditions.

In the following Sects.~\ref{s:PS} and \ref{s:Act} we  will find
explicit solutions for the elastic stress tensor in the presence
of turbulence which share a structure similar to \REF{R0eq}.
Namely, for $\De\gg 1$ the leading contribution to each component
of $\bm{\Pi}_0$ can be presented  as \REF{Pi}, in which  $C$,
given by Eqs.~\Ref{C} and \Ref{R2}, is some constant of the order
of unity, depending on the anisotropy of turbulent statistics.

\section{Polymer stretching in the passive regime}\label{s:PS}
\subsection{Cross-correlation tensor $\bm{Q}\Sb T$ in Gaussian turbulence}

When $\De\gg 1$ the characteristic decorrelation time of turbulent
fluctuations, $\tau\sb{cor}\lesssim 1/S_0$, is much smaller than
the polymer relaxation time $\tau\sb p$. Accordingly the turbulent
fluctuations can be  taken as $\delta$-correlated in time:
 \begin{equation}\label{Sigma}
    \<s^{ij}(t)s^{i'j'}(t')\> =
      \Xi^{ii',jj'}\sb{pas}\delta(t-t')\ .
    \end{equation}
This approximation is valid below the threshold of coil-stretched
transition, when the polymers do not affect the turbulent
statistics. This regime will be referred to  as the ``passive"
regime. The fourth-rank tensor $\bm\Xi\sb{pas}$ defined by
\REF{Sigma} is symmetric with respect to permutations of the two
first and two last indices,
$\Xi^{ii',jj'}\sb{pas}=\Xi^{i'i,j'j}\sb{pas}$. Incompressibility
leads to the restriction $\Xi^{ik,jk}\sb{pas}=0$. Homogeneity
implies
 $\Xi^{ii',jj'}\sb{pas}=\Xi^{ii',j'j}\sb{pas}$.

At this point we assume also Gaussianity of the turbulent
statistics. Then the tensor $\bm Q\Sb T$ can be found using the
Furutsu-Novikov decoupling procedure developed in
\cite{Novikov,Furutsu} for Gaussian processes: 
    \begin{equation}\label{Q-Gauss}
    Q^{ij}\Sb T = \Xi^{ij,kk'}\sb{pas}\Pi_0^{kk'}\ .
    \end{equation}
The cross-correlation tensor $ \bm{Q}\Sb T$ is proportional to a
presently undetermined stress tensor $\bm{\Pi}_0$. To find
$\bm{\Pi}_0$ one has to substitute \REF{Q-Gauss} into \REF{eqR0}
or into its formal solution \Ref{R0eq} and to solve the resulting
linear system of four equations for the non-zero components of
$\bm{\Pi}_0$:  $\Pi^{xx}$, $\Pi^{yy}$,  $\Pi^{zz}$ and
$\Pi^{xy}=\Pi^{yx}$.

The resulting equations are quite cumbersome. The basic physical
picture simplifies however in two limiting cases: i) isotropic
turbulence, Sec.~\ref{ss:PSb},  and ii)  anisotropic turbulence in
the limit of strong shear, Sec.~\ref{ss:PSc}.

\subsection{Isotropic turbulence} \label{ss:PSb}

First we consider the simplest case of isotropic turbulence. In
this case the tensor $\bm\Xi\sb{pas}$ has the form
\begin{eqnarray}\label{Gn}
\Xi^{ii',jj'} \sb{pas}= \Xi\[ \delta^{ii'}\delta^{jj'}
-\frac14\(\delta^{ij}\delta^{i'j'} +\delta^{ij'}\delta^{i'j} \)\]
\ , \label{Gn-b}\end{eqnarray} 
where $\Xi$ is a constant measuring the level of turbulent fluctuations.
Using \REF{Gn-b} in \REF{Q-Gauss} for  $\bm Q\Sb T$ one
gets
    \begin{equation}
    \bm Q \Sb T= \Xi\(\Pi_0\1 -\frac12\,\bm\Pi_0\)\ .
    \end{equation}
Substituting this relationship  into (\ref{eqR0a}) one gets a closed
equation for $\bm\Pi_0$:
\begin{subequations} \begin{eqnarray}\label{A6}
 \Big(\frac{D}{D\,t}+\frac{1}{\~
 \tau\sb p}\Big) \bm\Pi_0 \!\! &=&\!\! \S_0\*\bm\Pi_0 +\bm\Pi_0\*\S_0\Trans
\\&&\nn
+\(\Xi\Pi_0 +\frac{1}{\tau\sb p}\Pi\sb{eq}\)\1 \,, \\
\frac{1}{\~\tau\sb p} &\equiv& \frac{1}{\tau\sb p}
+\frac{\Xi}{2} \ . \label{A7}
\end{eqnarray}
\end{subequations}
The stationary solution of Eq.~(\ref{A6}) has the form
\begin{subequations}\label{A8}
\begin{eqnarray}\label{A8a}
\bm\Pi_0 &=& \Pi_0^{yy}\( \1 +\~\tau \sb p\(\S_0 +\S_0\Trans\)
+2\~\tau^{2}\sb p\S_0\*\S_0\Trans \)
    \,, \nn\\
    \Pi_0^{yy} &=& {\~\tau\sb p}\(\Xi\Pi_0+\frac{1}{\tau\sb p}
    \Pi\sb{eq}\)\,, \br
    \Pi_0 &\equiv& {\rm Tr}\,\{\bm\Pi_0\}\ .
\end{eqnarray}
In components:
\begin{eqnarray}\label{A8b}
\bm\Pi_0 &=& \Pi_0^{yy} \( 
\begin{array}{ccc}
2\,\~{\De}^2 +1 & ~~\~{\De}~~ & ~0~ \\
\~{\De}  & 1 & 0\\
0 & 0  & 1
\end{array}\)
    \,,\\\label{A8c}
    \~{\De} &\=& \~\tau\sb p \,S_0 < \De\ . 
\end{eqnarray}
\end{subequations}
We see that the tensor structure of $\bm\Pi_0$ has the
same form as in the laminar case, but with an
increased relaxation frequency given by Eq.~(\ref{A7}).

Taking the trace of Eq.~(\ref{A8}), one gets the following equation for
$\Pi_0^{yy}$:
\begin{eqnarray}\label{Gamma}
\Pi_0^{yy} &=& \frac{1}{\tau\sb p}\, \Gamma \,\Pi\sb{eq}
    \,, \\
 \Gamma & \equiv & \frac{1}{\tau\sb p} -\(2\,\~{\De}^2+\frac52\)\Xi \ .
\end{eqnarray}%
We see that the effective damping $\Gamma$ decreases with
increasing the turbulence level $\Xi$. At some critical value
of $\Xi=\Xi\sb{c}$, the effective damping goes to zero,
$\Gamma\to0$, and   formally $\Pi_0\to\infty$. The critical value
$\Xi\sb{c}$ corresponds to the threshold
 of the coil-stretch transition. Substituting \REF{A8c} into the
threshold condition $\Gamma=0$, one gets a 3rd-order
algebraic equation for $X\=\Xi\sb{c}\tau\sb p$:
    \begin{equation}
    5X^3 +18X^2 +4(3+4\,\De ^ 2)X = 8
    \ .\end{equation}
This equation has one real root $\Xi\sb{c}$, which we consider in two
limiting cases:
\begin{itemize}
\item Small shear $S_0\ll{1}/{\tau\sb p}$ ($\De\ll1$). In this
case the threshold
turbulence level $\Xi\sb{c}$ is proportional to the polymer
relaxation frequency
 \begin{subequations}\label{th}
\begin{eqnarray}
\Xi\sb{c} &\simeq& \frac{2}{5\, {\tau\sb p}}\(1 -\frac{5\,\De ^
2}{9}\) \,.\end{eqnarray} 
\item Large shear $S_0\gg {1}/{\tau\sb p}$ ($\De\gg1$). In this
case the threshold velocity gradient is much smaller.
Indeed:
\begin{eqnarray}\label{Sth-large}
\Xi\sb{c}  &\simeq& \frac1{2\,\tau\sb p\,\De ^ 2}\(
1-\frac{3}{4\,\De ^ 2}\) \,.\end{eqnarray}
\end{subequations}%
\end{itemize}

Evidently, the strong mean shear decreases the threshold of the
coil-stretch transition very significantly, by a factor of $\De ^
2$. The important conclusion is that in the entire region below
the threshold, $\Xi\le\Xi\sb{c}$, the renormalization of the
Deborah number can be safely neglected:
    \begin{equation}
        \De-\~{\De} \le \frac{1}{4\,\De} \ll 1
        \ .
    \end{equation}
This means that the structure of the elastic stress tensor in the
passive regime \Ref{A8b} hardly differs from the laminar case \Ref{R0eq}.

\subsection{The elastic stress tensor
 in anisotropic turbulent field}\label{ss:PSc}

Here we consider the case of strong shear, $\De \to\infty$, having
in mind that in the passive regime the tensorial structure of
$\bm\Xi\sb{pas}$ [see \REF{Gn-b}] is independent of $\De$. In this
case the leading contributions to each component of $\bm\Pi_0$ in
\REF{R0-sol-b} are:
    \begin{eqnarray} \nn 
\Pi^{xx}_0&=& 2\tau\sb p\,\De ^ 2\,Q^{yy}\,,\\\label{Pi0-lead} 
\Pi^{xy}_0&=&  \tau\sb p\De\, Q^{yy}\,, \br 
\Pi^{yy}_0&=& \tau\sb pQ^{yy}\,,  \qquad \Pi^{zz}_0= \tau\sb
pQ^{zz}\,,\br 
\Pi_0^{xz}&=&\Pi_0^{yz}= 0\ . 
\end{eqnarray} 
This means that for $\De\gg1$ $\bm\Pi_0$ takes on the form
\Ref{Pi} with
    \begin{equation}\label{C}
    C \simeq \frac{\Xi^{yy,xx}\sb{pas}}{\Xi^{zz,xx}\sb{pas}}\ .
    \end{equation}

The threshold condition for $\bm\Xi$ can be found along the
lines of the preceding subsection. The result is that only one
component of the tensor $\bm\Xi\sb{pas}$ is important:
\begin{equation} \Xi^{yy,xx}\sb{pas,c} \simeq
\frac 1{2\,\tau\sb p\,  \De ^ 2} \ .\end{equation} 
This result coincides to leading order with \REF{Sth-large} in
which $\Xi^{yy,xx}\sb{pas}\Rightarrow\Xi$ according to
\REF{Gn-b}.

Finally, we point out that the results presented in this section
remain valid for finite, but small, decorrelation time
$\tau\sb{cor}\to0$ in which $\Xi$ can be defined as
$\Xi\equiv^2\sigma_02\tau\sb{cor}/15$, where
$\sigma_0^2\equiv\<(\p^\alpha u^\beta)2\>={\rm Tr}\,\<\s\*\s\Sp
T\>$.

Notice that experimental and numerical
studies~\cite{75Vir,00SW,DNS1} show that the level of turbulent
activity in turbulent polymeric flows is of the same order as in
Newtonian flow at the same conditions. Simple estimations for the
typical conditions in  the MDR regime, when one observes large
drag reduction, show that the parameter $\Xi$ is far above the
threshold value \Ref{Sth-large}. Therefore, polymers in the MDR
regime cannot be considered as passive: there should be some
significant correlations between polymers and fluid motion that
prevent polymers from being infinitely extended despite of
supercritical level of turbulence. The character of these
correlations is  clarified in the following Section.


\section{Active regime of the polymer stretching}\label{s:Act}

In this Section we continue the discussion of the case of
homogeneous turbulence with mean shear, but allow for
supercritical levels of turbulent fluctuations, at which  the
polymers are strongly stretched. In that case they can no longer
be considered as a passive field that does not affect  the
turbulent fluctuations. In this \emph{active regime of the polymer
stretching} one cannot use \REF{Q-Gauss} for the cross-correlation
tensor $\bm{Q}\Sb T$ of the turbulent velocity field  $\B
v(\bm{r},t)$ and polymeric stretching $\B \pi(\bm{r},t)$. In the
present Section we reconsider the $\bm{v}-\bm{\pi}$ correlations
and show that in the active regime
 they   are determined by the so-called \emph{hydroelastic waves}
which are involved in transforming turbulent kinetic energy into
polymer potential energy and vice versa. Therefore in order to
find $\bm{Q}\Sb T$ in the active regime we need first to study the
basic properties of the hydroelastic waves themselves. This is
done in Sects. \ref{ss:Act-a} and \ref{ss:Act-b}. In
Sec.~\ref{ss:Act-a} we demonstrate that the coupled Eqs.~\Ref{19}
for $\B v$ and $\B \pi$ give rise to propagating hydro-elastic
plane waves
\begin{subequations}\label{DL}
    \begin{equation}\label{DL-c}
    \v_{\k}\,, \B b_\k \propto \exp [ i\(\k\* \r-\omega_\k t\)-\g_k t]
    \end{equation}
 with the dispersion law $\omega_\k$ and damping $\gamma_\k$:
    \begin{eqnarray}\label{DLa} 
\label{DLb} 
\omega_\k &=& \sqrt {(\k\* \bm{\Pi}_0\* \k)}\,, \
\\  
\g_k&=&\frac12\(\frac{1}{\tau\sb p}+ \nu_0 k^2\)\ .
\end{eqnarray} 
\end{subequations}
For large shear, $\De\gg 1$, these waves (with the exception of
those propagating exactly in the stream-wise direction $\x$) have
high quality-factor: $\omega_\k\gg \g _k$. The desired cross-correlation
tensor $\bm{Q}\Sb T$ is defined by the polarization of the
hydro-elastic  waves, that is the subject of
Sec.~\ref{ss:Act-b}.
In Sec.~\ref{ss:Act-c} we derive  the following equation
for tensor $\bm{Q}\Sb T$
\begin{equation}\label{act}
Q\Sb T^{ij}=\Xi^{ij,kl}\sb{act}\Pi^{kl}_0\,,
\end{equation}
that is very similar to the corresponding \REF{Q-Gauss} in the
passive regime. However  in \REF{act} the proportionality tensor
$\bm{\Xi}\sb{act}$, given by \REF{Q1}, is very different from
the corresponding tensor $\bm{\Xi}\sb{pas}$ in the passive
regime.

\subsection{Frequency and damping of hydro-elastic waves}\label{ss:Act-a}
The equations~\Ref{15} for the vector $\bm{v}$ and \emph{tensor}
$\bm\pi$ can be  reformulated in terms of  a new \emph{vector}:
\begin{subequations}\label{19}\begin{equation}\label{19-a}
\B b\equiv \tau\sb p\bn\*\bm\pi\,, \end{equation}
instead of the tensor $\bm{\pi}$.  The new equations read
\begin{eqnarray}\mkern-50mu
\label{19b}\frac{D\v }{D\, t} &=& -\S_0\*\v +\nu_0\Delta\v -\bn p
+  \g \sb p\, \B b +\BC N_v\,, \\ \label{19c}\mkern-50mu
\frac{D \B b}{D\, t} &=& -\frac{1}{\tau\sb p} \B b + \S_0\*\B b
-\tau\sb p\^ \Omega ^ 2\,\v +\BC N_b\,,\end{eqnarray} 
where
 \begin{eqnarray}
 {\^\Omega}^2  &\equiv&
 -\bm\nabla\cdot\bm\Pi_0\cdot\bm\nabla\,,\\
    \BC N_v &\=& -\bn\*\v\v
    \,,\quad
    \BC N_b \= \tau\sb p\bn\*\bm\Phi_\pi\ .
\end{eqnarray}
\end{subequations}
The linearized version of \Ref{19}  gives rise to hydro-elastic
waves \cite{steinberg,01BFL} with an alternating exchange between the
kinetic energy of the carrier fluid and the  potential
energy of the polymeric subsystem. In the simplest case of
space-homogeneous turbulent media (no mean shear, $S_0=0$)  the
homogeneous Eqs.~\Ref{19} have plane wave solutions cf. Eq. \Ref{DL}.

When the mean shear exists it appears in Eqs. ~\Ref{19}  with a
characteristic frequency  $S_0$. In the region
of parameters that we are interested in,  $S_0$ is much smaller
than the wave frequency $\omega_\k$, but  much larger than the
wave damping frequency $\g_k$.  This means that the shear does not
affect the wave character of the motion, but it can change the
effective damping of the plain wave with a given wave vector $\k$.
Indeed, due to the linear inhomogeneity of the mean velocity
(constant shear) the wave vector becomes time dependent according
to
\begin{equation}\label{kt} 
\frac{d \k }{d t}= - \S _0\Trans \* \k\ . \end{equation}
This means that in the case $S_0\gg \g_k$ the shear frequency
$S_0$ serves as an effective de-correlation frequency instead of
$\g_k$: \begin{equation}\label{estg} \gamma_{\bm{k}}\Rightarrow
\gamma = b S_0\,,
\end{equation}%
where $b$ is the dimensionless constant of the order of
unity.

We reiterate that the limit $\De\gg 1$ is consistent with the
frequency of the hydro-elastic waves $\omega_\k$ being much larger
than their effective damping.

\subsection{Polarization of hydro-elastic waves}\label{ss:Act-b}

The formal solution of the linear \REF{19c} for $\B b$ can be written in
terms of the Green's function $\^G\sb p$:
\begin{subequations} \begin{eqnarray}\label{b-a}
\B b&=& -\tau\sb p\(\^G\sb p +\S_0\*\^G\sb p^ 2\)
\^\Omega ^ 2 \, \v\,,  \\
\^G\sb p &\equiv& \(\frac{D}{D\, t} +\frac{1}{\tau\sb p}\)^{-1} \
.
\end{eqnarray}\end{subequations} 
Denoting by $F_n(t,\r)$ the result of $n$-fold action $\^ G\sb
p^n$ on the function $f(t,\r)$, 
\begin{subequations}\label{bla}
\begin{equation} F_n(t,\r)\equiv \^ G\sb p^n f(t,\r)\,,
\end{equation}
 one writes
 \begin{eqnarray}
F_1 (t,\r )&=& \int\limits _0^{\infty}d\tau f(t-\tau,
\r-\S_0\*\r\tau)\exp \(-\frac{ \tau}{\tau\sb p} \) \,,\nonumber\\
\br
 F_n (t,\r ) &=& \int\limits
_0^{\infty}\frac{d\tau \, \tau^{n-1}}{(n-1)!} f(t-\tau,
\r-\S_0\*\r\tau)\exp\(-\frac{ \tau }{\tau\sb p}\)
\,.\end{eqnarray}
\end{subequations}

By straightforward calculations one can show that $\^G\sb p$
satisfies the following  commutative relationship
\begin{equation}\label{bla}\^G\sb p\bn - \bn\^G\sb p= \S_0\Trans\*\^G\sb p^
2\bn\ .
\end{equation}%
Using this relation in \REF{b-a} repeatedly we can rewrite $\B b$
as
\begin{eqnarray}\label{pol}
\B b&=& \tau\sb p \[ \(\^G\sb p\bn\*{\bm{\Pi}}_0\*\bn\)
+\S_0\*\(\^G\sb p^ 2\bn\*{\bm{\Pi}}_0\*\bn\)\] \v \nonumber\\&=&
 \tau\sb p \Big[ \(\bn\*{\bm{\Pi}}_0\*\^G\sb p\bn\)
 +\(\bn\*\S_0\*{\bm{\Pi}}_0\*\^G\sb p^ 2\bn\) \nonumber\\&&
 +\S_0\*\(\bn\*{\bm{\Pi}}_0\*\^G\sb p^ 2\bn\)
\\\nonumber&&
+2\S_0\*\(\bn\*\S_0\*{\bm{\Pi}}_0\*\^G\sb p^3\bn\) \Big] \v
\,.\end{eqnarray}
This  is an exact relationship for the  polarization of the
hydro-elastic waves in the presence of mean shear, and it will be
used in the  analysis of the cross-correlation tensor $\bm{Q}\Sb
T$.

\subsection{General structure of the cross-correlation tensor}
\label{ss:Act-c}
 The cross-correlation tensor $\bm{Q}\Sb T$ \REF{eqR0c}  can be
rewritten in terms of the vector field $ \B b$  as follows:
\begin{equation}\label{Qb} \bm{Q}\Sb T =  -
\frac{1}{\tau\sb p}\< \v \B b + \B b \v \>\ .
\end{equation}
 Substituting the vector $\B b$ from \REF{pol} into \REF{Qb}  one
 finds that the cross-correlation tensor $\bm{Q}\Sb T$ is
 proportional to the elastic stress tensor $\bm{\Pi}_0$,  according
 to \REF{act}, in which the proportionality tensor $\bm{\Xi}\sb{act}$
is expressed  in terms of the second-order correlation functions
of the velocity gradients, as follows:
\begin{subequations}\label{Sijkl}
\begin{eqnarray}\nn
\Xi^{ijkl}\sb{act} &=&\Xi_1^{ijkl} +\Xi_1^{jikl}
+\(\Xi_2^{ijk'l} +\Xi_2^{jik'l}\) S_0^{k'k}
\\  \label{Q1}&&
+\( S_0^{jj'}\Xi_2^{ij'kl} +S_0^{ii'}\Xi_2^{ji'kl} \)
\\\nonumber&&
+2\( S_0^{jj'}\Xi_3^{ij'k'l} +S_0^{ii'}\Xi_3^{ji'k'l}
\)S_0^{k'k} \ .
\end{eqnarray}
Here $\bm{\Xi}_n$ are time-integrated tensors defined as
follows:
\begin{eqnarray}\label{Sijkl-b}
 \bm\Xi_n \equiv
\int_0^\infty \frac{d\tau \, \tau^{n-1}}{(n-1)!}\, \B
\Xi(\tau)\exp\( -\frac{\tau}{\tau\sb p} \) \,\\
\label{Sijkl-a} \Xi^{ijkl}(\tau) \equiv \< \frac{\p v^i(t,
\r)}{\p r^k} \frac{\p v^j(t-\tau, \r')}{\p r^{'l}} \>\
\end{eqnarray}
\end{subequations}
and  $\r' = \r -\tau\, \S_0\*\r$.

In order to estimate the relative importance of $\Xi_1$, $\Xi_2$,
and $\Xi_3$ in \REF{Q1} notice that the integrals in \REF{Sijkl-b}
are  dominated by contributions from  the longest hydro-elastic
waves in the system. These longest $k$-vectors have a
decorrelation time $\gamma$ [which is estimated in Eq.
(\ref{estg})], and are characterized by a frequency $\omega$ that
we will be  specified below. Using these estimates we can
approximate the time dependence of $\Xi^{ijkl}(\tau)$  as follows
\begin{subequations}\begin{eqnarray}\label{fac1a} 
\Xi^{ijkl}(\tau) &\Rightarrow& \sigma^{ijkl}f(\tau)\,,
\quad
  \bm\sigma \equiv \bm\Xi(0)
 \\
\label{fac1b} f(\tau)&=& \cos \( \omega  \, \t\) \,
 \exp\( - \gamma \t \)\ .
\end{eqnarray}\end{subequations}
Note that the assumption here is that $\omega$ and $\gamma$ do not
depend on the tensor indices. This is a simplifying assumption
that does not carry heavy consequences for the qualitative
analysis.

Under these assumptions the tensors $\B \Xi_n$ in
\REF{Q1} can be estimated as follows
\begin{subequations} \label{fn}\begin{eqnarray}
\Xi^{ijkl}_n &=&  f_n \sigma^{ijkl}\,,\br f_n &=&\int_0^\infty
\frac{d\tau \, \tau^{n-1}}{(n-1)!}\,
 f(\t) \exp\( -\frac{\tau}{\tau\sb p} \)\\
 &\simeq& \frac{\text{Re}(\g+i\omega)^n}{\(\g^2+\Omega ^ 2\)^n}
 \,,\\
 f_1&\simeq&  \frac {\g}{\Omega ^ 2}\,, \quad
f_2\simeq  - \frac {1}{\Omega ^ 2} \,,\quad  f_3\simeq  - \frac
{3\, \gamma}{\omega ^ 4 }\ .
\end{eqnarray}\end{subequations} 
At this point we need to estimate $\omega$. From Eq. \Ref{DLb} we
see that we need to take the largest of component of $\bm{\Pi}_0$,
and smallest available $k$ vector which will be denoted as $k_{\rm
min}$. Since we expect the $xx$ component of $\bm{\Pi}_0$ to be
the largest component we write
\begin{equation}
\label{esto}
\omega\simeq \sqrt{\Pi^{xx}_0}\,  k_{\rm min} \ .
\end{equation}%
We should note at this point that homogeneous turbulence has no
inherent minimal $k$ vector, since there is no natural scale. In
reality there is always an outer scale which is determined by
external constraints. For further progress in the analysis of the
structure of the cross-correlation tensor one need to specify the
outer scale of turbulence.  To this end we will consider in the
next Section homogeneous turbulence with a constant shear as an
approximation to wall bounded turbulence of polymeric solution in
the region of logarithmic mean velocity profile.

\section{Polymer stretching  in wall bounded
turbulence}\label{s:Disc}

In this section we consider turbulent polymeric solutions in wall
bounded flows, and show that the elastic stress tensor takes on
the universal form \Ref{Pi}. This implies very specific dependence
of the components of the elastic stress tensor on the distance
from the wall. The theoretical predictions will be checked against
numerical simulations and will be shown to be very well
corroborated. Since we are interested in drag reduction we must
consider here the active regime when the polymers are sufficiently
stretched to affect the turbulent field. As mentioned  before,
large drag reduction necessarily implies $\De\gg 1$. For
concreteness we restrict ourselves by considering the most
interesting logarithmic-law region. Extension of our results to the
entire turbulent boundary layer is straightforward.

\subsection{Cross-correlation tensor  in wall bounded turbulence}
In this section we consider in more details tensor $\bm{Q}\Sb T$
for wall bounded turbulent flows. In this case  the outer scale of
turbulence is estimated as the distance to the wall. Therefore in
Eq. ~\Ref{esto} $k_{\rm min}\sim 1/y$ where $y$ is the distance to
the wall.

Also we can use the fact that in the turbulent boundary layer the
mean velocity has a logarithmic profile for both Newtonian and
viscoelastic flows, (with slopes that differ by approximately a
factor of five). Therefore  the  mean shear $S_0$ is inversely
proportional to the distance to the wall $y$ and can be estimated
as
 \begin{equation}\label{New} 
S_0\simeq  \frac{\sqrt{\mathcal{P}}}{y}\,,
    \end{equation}
where $\mathcal{P}$ is the total flux of the mechanical momentum
 (for example, in the channel of half-width $L$, equal to $p'\,L$,
where $p'$ is the pressure gradient in the streamwise direction).

Another well established fact is  that when the effect of drag
reduction is large, the momentum flux toward the wall is carried
mainly by the polymers. Then, \REF{bal1} gives
\begin{equation}\label{est1} \Pi^{xy}_0\simeq \mathcal{P}\simeq
(S_0\, y)^2\ .
    \end{equation}

To estimate the frequency (\ref{esto}) we need to handle the other
component of the elastic stress tensor. Examining Eqs.
(\ref{R0-sol-b}) in the limit $\De\gg 1$ we will make the
assumption that the inequalities (\ref{case}) hold also in the
strongly active regime. This assumption will be justified
self-consistently below. It then follows immediately that
\begin{equation}\label{est2}
\Pi_0\simeq \Pi_0^{xx}\simeq  \De (S_0\, y)^2\,, \quad
\Pi_0^{yy}\simeq \Pi_0^{zz}\simeq \frac{(S_0\, y)^2}{\De}\ .
    \end{equation}
Now we can estimate the characteristic frequency of
hydro-elastic waves in \REF{esto} as follows: 
\begin{equation}\label{est-o}
\omega = a\, \sqrt{\mathcal{D}e}\, S_0\,,
\end{equation}%
where the dimensionless parameter $a\simeq 1$.  One sees that $\omega$
indeed is much larger than $\gamma=b\,S_0$, \REF{estg}, as we
expected.

Next we can continue the analysis of the cross-correlation tensor
in  wall bounded turbulence.  Using the estimates~ \Ref{estg},
\Ref{fn} and \Ref{est-o}, we notice that
$$
\Xi_1 \simeq S_0 \Xi_2  \simeq S_0^2 \mathcal{D}e \Xi_3 \gg S_0^2
\Xi _3\ .
$$
Therefore one can neglect terms with $\Xi_3$ on  the RHS of
\REF{Q1}. Moreover, we can further simplify \REF{Q1}, taking into
account that the solution of the system  of Eqs. \Ref{eqR0} and
\Ref{Q1} preserves the structure of polymer stress tensor
\Ref{R0-sol-b} in which $\Pi_0^{xx}\gg\Pi_0^{xy}$. This allows one
to neglect the  two last terms in the first line of the
RHS of \REF{Q1}, where only $S^{xy} \neq 0$. After that
the cross-correlation tensor $\bm{Q}\Sb T$, given by \REF{act} in
terms of tensor $\bm{\Xi}\sb{act}$, \REF{Q1}, can be
represented via the second order tensor 
\begin{subequations}\label{B} \begin{equation}\label{Ba} 
B^{ij}\equiv \sigma^{ij,kl}\Pi_0^{kl} \end{equation} 
as follows: 
\begin{equation} \label{Bb} 
\bm{Q}\Sb T= \frac1{a^2 \De S_0 }\[ b \(\B B + \B B\Trans\)- (\B
B\*\y)\x -\x(\B B\*\y)\] \ .
\end{equation} \end{subequations}

Recall that we are looking for the tensorial structure of the
cross-correlation tensor $\bm{Q}\Sb T$ in order to find the
structure of $\bm{\Pi}_0$. One sees from \REF{Ba} that although
the tensor $\bm{\Pi}_0$ is strongly anisotropic (its components
differ by powers of $\De$), once contracted with the tensor
$\sigma^{ij,kl}$  of a general form, it   gives a tensor $\bm{B}$
with components of the same order in $\De$.
  Then \REF{Bb} implies that the cross-correlation tensor
$\bm{Q}\Sb T$ has components that are all of the same order in $\De$.
This result is even stronger than the assumption \Ref{case} in our derivation
 (that we needed to recapture self-consistently). Having done so we
can conclude that the elastic stress tensor $\bm{\Pi}_0$ has the
universal tensorial structure given by \REF{Pi}.

Armed with this knowledge we observe that  the leading
contribution to $\B B$ on the RHS of \REF{Ba} is given by
$\Pi_0^{xx}= 2\,\De ^ 2 \Pi_0^{yy}$: 
\begin{subequations}\label{B1}\begin{equation} \label{B1a}
 B^{ij}= \sigma^{ij,xx}\Pi_0^{xx}=2\, \De ^ 2\,
 \sigma^{ij,xx}\Pi_0^{yy}\ . 
\end{equation}
According to definition \Ref{Sijkl-a} the tensor
$\sigma^{ij,xx}=\Xi^{ij,xx}(0)$ can be evaluated as $c
W^{ij}/y^2$, where $y$ is the outer scale of turbulence (and
distance to the wall), the Reynolds stress tensor $\B W$ was
defined by \REF{RS} and a new dimensionless constant $c$ is of the
order of unity. Thus one has
\begin{equation}\label{B1b}
 B^{ij}=2\,c\,\De ^ 2\, W^{ij}\Pi_0^{yy}\Big / y^2 \ .
\end{equation}  \end{subequations} 
Now we can write an explicit equation for $\bm{Q}\Sb T$:
\begin{equation}\label{Q11} 
\bm{Q}\Sb T= d\, \frac{\tau\sb p}{y ^2}\, \Pi _0^{yy}\,
\left\{\B W- \frac{1}{2b}\[ \(\B W\* \y\)\, \x +\x \(\B W \*
\y\)\] \right\} \,, 
 \end{equation}
where $d\equiv 4b\,c/a^2$.
\subsection{Explicit solution for the polymeric stress
tensor}\label{ss:explicit}
Having \REF{Q11} for $\bm{Q}\Sb{T}$ in terms of
$\bm{\Pi}_0$ and $\bm{W}$ we can find an explicit solution
 for the mean elastic stress tensor $\bm{\Pi}_0$ in the presence of
intensive turbulent velocity fluctuations and strong shear. As a
first step in \REF{eqR0b} for $\bm{Q}$ we neglect the equilibrium
term $\Pi\sb{eq}\1/{\tau\sb p} $ since it is expected to be much
smaller than  $\bm{Q}\Sb T$, which stems from turbulent
interactions. Then we substitute $\bm{Q}=\bm{Q}\Sb T$ into
\REF{Ro-a} [or into \REF{R0-sol-a}] and solve the resulting
equations.   To leading order in $\De=S_0\tau\sb p\gg 1$ (in each
component) this solution  takes the form~\REF{Pi} with
 \begin{equation}\label{R2}
   C=  W^{zz}/W^{yy} \ .
 \end{equation}
Notice that  in our approach the ``constant" shear has to be
understood as a local, $y$-dependent shear in the turbulent channel
flow, according to \REF{New}. Correspondingly, the Deborah number
also becomes $y$-dependent.
 \begin{equation}\label{de1}
 \De\Rightarrow \De(y)= S_0(y) \tau \sb p \simeq \frac{\sqrt {\mathcal{P}}\tau \sb p  }
 y\ .
\end{equation}%
Now Eqs.~\Ref{est1} and \Ref{est2} provide an explicit dependence of the
components of tensor $\bm{\Pi}_0$ on the distance from the wall:
\begin{eqnarray}\label{pi2}\nn
    \Pi_0&\simeq&  \Pi_0^{xx}\simeq \frac{\mathcal{P}^{3/2}\tau\sb
    p}y\,, \\
 \Pi_0^{xy}&=& \Pi_0^{yx}  \simeq \mathcal{P}\,, \br
\Pi_0^{yy}&\simeq&  \Pi _0^{zz}\simeq \frac{\sqrt {\mathcal{P}}\,
y}{\tau \sb p}\ .
    \end{eqnarray}%

At this point we can summarize our procedure as follows. First, we
\emph{assumed} that the $\bm{Q}$ tensor is not strongly
anisotropic such that weak inequalities~\Ref{case} are valid. This
allowed us to use the universal form~\Ref{Pi} of tensor $\bm{\Pi}_0$
in actual calculations of d the cross-correlation function
$\bm{Q}\Sb T$.  Then we demonstrated that  $\bm{Q}\Sb T$ \emph{
indeed satisfies the required inequalities}~\Ref{case}. This means
that \REF{Pi} presents \emph{a self-consistent solution of the
exact equations} in the limit $\De \to \infty$.  The described
procedure does not guarantee that ~\REF{Pi} is \emph{the only
solution} of the problem at hand. However we propose that this
solution is the realized one, and we will check it next against
numerical simulations.

\subsection{Effective polymeric viscosity}\label{ss:Disc}

Armed with the structure~\REF{Pi} of the polymeric
 stress tensor, we can  rewrite the   equation of the mechanical
 balance~\Ref{bal1}
   as follows:
\begin{equation}\label{flux1}
{\cal P}= \nu_0S_0 -W^{xy} +\tau\sb p\Pi_0^{yy}S_0\ .
\end{equation} 
This means, that the polymeric contribution to the momentum flux (last
term on the RHS of this equation) can be considered as an
``effective polymeric viscosity" $\nu\sb p$ 
   \begin{subequations}\label{est3}
   \begin{equation}\label{est3-a}
\nu_0\Rightarrow \nu_0+\nu\sb p\,, \quad     \nu\sb p = \tau\sb
p\Pi_0^{yy}\ .
    \end{equation}
Using \REF{pi2} one gets universal ($\tau\sb p$-independent)
linear dependence of $\nu\sb p$: \begin{equation}\label{lin}
\nu\sb p \simeq \sqrt{\mathcal{P}}\, y\,,
\end{equation}\end{subequations}
as was suggested in ~\cite{03LPPT} (and see also \cite{04BDLPT}).

The polymeric contribution to the rate of turbulent energy
dissipation has the form:
\begin{equation}\label{ex1-a} {\ve\Sb T}\sb p= \text{Tr}\<\s \* \bm{\Pi}\>=\frac12
    \text{Tr}\{\bm{Q}\Sb T\}\,,
\end{equation}%
see  \REF{ex} in Appendix~\ref{ss:BEMc}. Using here \REF{Q11} one
gets\begin{equation}\label{ex1}  {\ve\Sb T}\sb p= \frac {d\,
\tau\sb p}{ 2\, y ^2} \, \Pi_0^{yy}
    \[ 2 K -\frac1b W^{xy}
    \]  \ ,\end{equation}
where $K=\frac12 { \rm Tr}\{\bm{W}\}$ is the density of the
turbulent kinetic energy.  Having in mind  that in the MDR regime
$W^{xy}\simeq K $, having the same dependence on the distance from
the wall, \REF{ex1} can be rewritten in terms  of the effective
polymeric viscosity $\nu\sb p$, given by \REF{est3}:
    \begin{equation}\label{est4}
 {\ve\Sb T}\sb p = \nu \sb p \< |\nabla \B u|^2\>,
    \end{equation}
where  $  \< |\nabla \B u|^2\> $ was estimated  as $K /y^2$.

Notice that the naive estimate for the effective polymeric viscosity
is $\~ \nu\sb p\simeq \tau\sb p\Pi_0$, that exceeds our
result~\Ref{est3} by a factor of $\De^2$. The reason for this difference
is the wave character of the fluid motion; the naive result is
valid for the estimate of the characteristic {\em instantaneous}
energy flux. As usual in high-quality waves or oscillations, the rate
of {\em energy exchange} between subsystems is much larger than
the rate of {\em energy dissipation}.

\subsection{Comparison with DNS data}

The tensorial structure of the polymer stress tensor  was studied
in fair detail in DNS for channel flows in various papers,
see~\cite{DNS1,Beta,00ACP,DNS2,DNS3} and references therein. The
main problem is that the large \Re\ regime at which the universal
MDR ~\cite{75Vir} is observed is hardly available. In these DNS the
maximal available $\De$ was below 100 at the wall, decreasing to
about 10 in the turbulent sublayer.  For these conditions  only up
to 50-60 \% of the total momentum flux is carried by the polymers.
Nevertheless we can compare our analytical results for $\De \to
\infty$ with DNS at moderate $\De$,  at least on a qualitative
level.

\begin{figure}
\includegraphics[width=0.5 \textwidth]{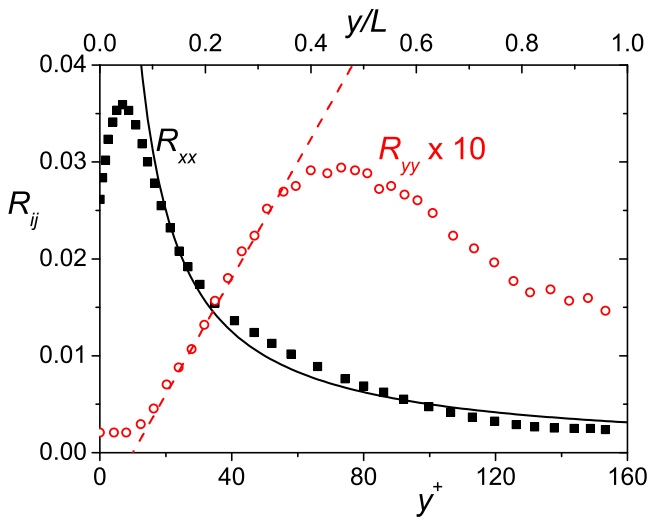}
  \caption{\label{f:1} Comparison of the DNS data Ref.~\cite{DNS2} for
the mean
  profiles of the $xx$ and $yy$ components of the elastic stress
  tensor $\Pi_0^{ij}$ with analytical predictions. Black squares:
  DNS data  for streamwise diagonal component $\Pi_0^{xx}$, that
  according to our understanding, \REF{pi2}, has to decrease as
  $1/y$ with the distance to the wall. Black solid line: $1/y^+$
  dependence. Red empty circles: DNS data for the wall-normal
  component $\Pi_0^{yy}$, for which we predicted linear increase
  with $y^+$ in the log-law turbulent region. Red dashed line -
  linear dependence, $\propto y^+$. }
\end{figure}

The most frequently studied object is $\Pi_0\equiv \mbox{Tr}
\{\bm{\Pi}_0\}$. This object is dominated by the stream-wise
component $\Pi^{xx}_0$, see, Fig.~12 in Ref.~\cite{DNS1},
Fig.~5.3.1 in Ref.~\cite{Beta}, Fig.~6 in~Ref.~\cite{DNS3}, etc.
The accepted result is that $\Pi_0$ decreases in the turbulent
layer. Our Eqs. \Ref{pi2} predict $\Pi_0\propto 1/y$  in this
region. This  rationalizes the DNS observations in
    Refs.~\cite{DNS1,Beta,DNS2,DNS3}.  As an
example, we present in Fig.~\ref{f:1} by black squares the DNS
data of~\cite{DNS2} for the largest, streamwise component
$\Pi_0^{xx}$ and plotted by solid black line the expected [see
\REF{pi2}] profile $\propto 1/y^+$. In spite of the fact that the
MDR asymptote with the logarithmic mean velocity profile is hardly
seen in that DNS, the agreement between the DNS data and our
prediction is obvious.

The effective polymeric viscosity was measured, for example, in
Ref.~\cite{DNS1}, Fig.~5, and in Ref.~\cite{Beta}. In these Refs.
$\nu\sb p (y)$ was understood as $\Delta\sb p(y)/S_0(y)$, where
$\Delta\sb p(y)$ is the Reynolds stress deficit, which is
$\Pi_0^{xy}$ according to \REF{bal1}. The observations of
Refs.~\cite{DNS1,Beta} is that $\nu\sb p (y)$ in the turbulent
boundary layer grows linearly with the distance from the wall.
Recent data of DNS of the Navier-Stokes equation with polymeric
additives (using the FENE-P model) taken from Ref.~\cite{DNS2} are
presented in Fig.~\ref{f:1}. The wall-normal component
$\Pi_0^{yy}$ is presented by the red empty circles. One sees that
(in the quite narrow) region $20<y^+< 60$ (where the mean velocity
profile is close of MDR) the profile $\Pi_0^{yy}$ is indeed
proportional to $y^+$, as represented by the red dashed line. This
is in agreement with our result, see \REF{lin} and our
Ref.~\cite{03LPPT}.

Direct DNS of other components of the polymeric stress tensor,
see, e.g. Fig.~6 in Ref.~\cite{DNS2} also agrees with the
tensorial structure of $\bm{\Pi}_0$ given by \REF{pi2}.

\section{Summary and discussion}\label{s:sum}

The aim of this paper was the analysis of the polymer stress
tensor in turbulent flows of dilute polymeric solutions. On the
face of it this object is very hard to pinpoint analytically,
being sensitive to complex interactions between the polymer
molecules and the turbulent motions. We showed nevertheless that
in the limit of very high Deborah numbers, $\De \gg 1$, this
tensor attains a universal form. Increasing the complexity of our
turbulent ensemble, from laminar, through homogeneous turbulence
with a constant shear, and ending up with wall bounded anisotropic
turbulence, we proposed a universal form

\begin{equation}
    \bm\Pi_0(y) \simeq \Pi^{yy}_0(y)\(\begin{array}{ccc}
        2\De ^ 2 (y)& ~\De(y)~ &  ~~0~~\\
        \De(y) & 1 & 0 \\
        0 & 0 & C(y)
    \end{array}\)\ .\label{R0turb-c}
    \end{equation}

We rewrite this form here to stress that it remains unchanged even
when the Deborah number, and with it the components of the tensor,
become space dependent. Obviously, this strong result is expected
to hold only as long as the mean properties, including the mean
shear, vary in space in a controlled fashion, as for example in
the logarithmic layer near the wall (be it a K\'arm\'an or a Virk
logarithm).

As an important application of these results we considered in
Sect. \ref{s:Disc} the important problem of drag reduction by
polymers in wall bounded flows. A difficult issue that caused a
substantial confusion is the relation between the polymer physics,
the effective viscosity that is due to polymer stretching, and
drag reduction. In recent work on drag reduction it was shown that
the Virk logarithmic Maximum Drag Reduction asymptote is
consistent with a linearly increasing (with $y$) effective
viscosity due to polymer stretching. This result seemed counter
intuitive since numerical simulations indicated that polymer
stretching is decreasing as a function of $y$. The present results
provide a complete understanding of this conundrum. ``Polymer
stretching" is dominated by $\Pi_0^{xx}$ since it is much
larger than all the other components of the stress tensor.
As shown above, this component is indeed decreasing when $y$
increases, cf. Fig. \ref{f:1}. On the other hand the effective
viscosity is proportional  $\Pi_0^{yy}$, and this component is
indeed increasing (linearly) with $y$, cf. Fig. \ref{f:1}. In
fact, drag reduction saturates precisely when $\Pi_0^{xx}$ and
$\Pi_0^{yy}$ become of the same order.

\begin{acknowledgements}
We thank Roberto Benzi for useful discussions. This work had been
supported in part  by the US-Israel Binational Science Foundation,
by the European Commission via a TMR grant, and by the Minerva
Foundation, Munich, Germany.
\end{acknowledgements}
\appendix
\section{Exact energy balance equations}\label{ss:BEMc}
In this Appendix we present exact  energy balance equations
that are useful in the analysis of turbulence of the polymeric
solutions. In the present study we employ only one of them,
i.e. \REF{ex}.

Introduce the mean densities of the turbulent kinetic
energy $E\Sb T$ and polymeric potential energy $E\sb p$
\begin{subequations}
\begin{eqnarray}
E \Sb T& \equiv & \frac12 W\,, \quad W\= \text{Tr}\,\{\B W\}\,,\\
E \sb p& \equiv & \frac12\Pi_0\,, \quad \Pi_0\={\rm
Tr}\,\{\bm\Pi_0\} \ .
\end{eqnarray}
\end{subequations}

Using Eqs.~\Ref{NSE} one can derive equations for the balance of
$E\Sb T$ and $E\sb p$ and for their sum:
\begin{subequations}\label{EB}
\begin{eqnarray}\label{EBa}
\frac{D E\Sb T}{D  t}   &=&\ve^+\Sb T - \ve^-\Sb T -{\ve\Sb T}\sb
p =0 \,, \\ \label{EBb} 
\frac{D E \sb p}{D t}  &=&\ve^+\sb p - \ve^-\sb p  + {\ve\Sb T}\sb
p=0\,,
\\ \label{EBc}
 \hskip -1cm \frac{D}{D t}\(E\Sb T+E \sb p \) &=& \ve^+\Sb T+
\ve^+\sb p - \ve^-\Sb T  - \ve^-\sb p=0 \ .
\end{eqnarray}
 \end{subequations}
\begin{subequations}
We denote by $\ve^+\Sb T$ and $\ve^+\sb p$  the energy flux from
the mean flow to the turbulent and polymeric subsystem
respectively:
 \begin{eqnarray}
\ve^+\Sb T &\=& - S_0 W^{xy}\,,\\
\ve^+\sb p &\=& S_0 \Pi_0^{xy}\,;
 \end{eqnarray}
$\ve^-\Sb T$ describes the dissipation of energy in the turbulent
subsystem, whereas $\ve^-\sb p$  is the dissipation in the
polymeric subsystem due to the relaxation of the stretched
polymers back to equilibrium:
\begin{eqnarray}
\ve^-\Sb T &\=& \nu_0 \text{Tr}\< \s \* \s \Sp T  \>\,,\\
\ve^-\sb p &\=&  \frac{1}{2\,\tau\sb p}{\rm Tr}\(\bm\Pi_0
-\bm\Pi\sb{eq}\)\ . \end{eqnarray} 
The last term on the RHS of Eqs. \Ref{EBa} and \Ref{EBb} [that is
absent in \Ref{EBc}]  describes the energy exchange between the
polymeric and turbulent subsystems:
\begin{equation}\label{ex} 
{\ve\Sb T}\sb p\equiv  \text{Tr}\<\s \* \B\pi\> = \frac12{\rm
Tr}\,\{\bm Q\Sb T\}\ .\end{equation}
 \end{subequations}
 Using the expression for the momentum flux, we obtain an exact
balance equation for the total energy
\begin{subequations}
\begin{equation} \label{mean-en}
E=E\Sb{V}+E\Sb T + E\sb p\,, 
\end{equation} that  in the stationary state reads:
\begin{equation}\label{tot-bal}
S_0{\cal P}= \frac{1}{2\tau\sb p}{\rm Tr}\(\bm\Pi_0
-\bm\Pi\sb{eq}\)
 +\nu_0\(S_0^2 +\Tr\<\s\*\s\Trans\>\)
\,.\end{equation}
\end{subequations}
In \REF{mean-en} $E\Sb{V}$ is the density of the kinetic energy of
the mean flow, defined up to an arbitrary constant, depending on
the choice of the  origin of coordinates.
 The LHS of  \REF{tot-bal} describes the work of external
forces needed to maintain the constant mean shear. The first term
on the RHS ($\propto {1}/{\tau\sb p} $) describes the energy
dissipation in the polymeric subsystem. The term $\nu_0 S_0^2$
represents the viscous dissipation due to the mean shear, while
the last term on the RHS is responsible for the viscous
dissipation caused by the turbulent fluctuations.

\end{document}